\documentclass[prl,superscriptaddress,nofootinbib,notitlepage,twocolumn,tightenlines]{revtex4-1}
\pdfoutput=1
\usepackage{graphicx,amsmath,bm,color}
\usepackage[pass]{geometry}
\usepackage[colorlinks = true, linkcolor = black, urlcolor  = blue, citecolor = blue, anchorcolor = blue]{hyperref}

\begin{document}
\title{\ \\[1em] \Large Coherent Dynamics Enhanced by Uncorrelated Noise}
\author{Zachary G. Nicolaou}
\affiliation{Department of Physics and Astronomy, Northwestern University, Evanston, Illinois 60208, USA}
\author{Michael Sebek}
\affiliation{Department of Chemistry, Saint Louis University, St. Louis, Missouri 63103, USA}
\affiliation{Network Science Institute, Northeastern University, Boston, Massachusetts, 02115, USA}
\author{Istv\'{a}n Z. Kiss}
\affiliation{Department of Chemistry, Saint Louis University, St. Louis, Missouri 63103, USA}
\author{Adilson E. Motter}
\email[email: ]{motter@northwestern.edu}
\affiliation{Department of Physics and Astronomy, Northwestern University, Evanston, Illinois 60208, USA}
\affiliation{Northwestern Institute on Complex Systems, Northwestern University, Evanston, Illinois 60208, USA}

\begin{abstract}
\begin{center} (Received 8 March 2020; revised 17 June 2020; accepted 21 July 2020)\\[1em]\end{center}

\small Synchronization is a widespread phenomenon observed in physical, biological, and social networks, which persists even under the influence of strong noise. 
Previous research on oscillators subject to common noise has shown that noise can actually facilitate synchronization, as correlations in the dynamics can be inherited from the noise itself. 
However, in many spatially distributed networks, such as the mammalian circadian system, the noise that different oscillators experience can be effectively uncorrelated. 
Here, we show that uncorrelated noise can in fact enhance synchronization when the oscillators are coupled. Strikingly, our analysis also shows that  uncorrelated noise can be more effective than common noise in enhancing synchronization.   
We first establish these results theoretically for phase and phase-amplitude oscillators subject to either or both additive and multiplicative noise. 
We then confirm the predictions through experiments on coupled electrochemical oscillators.  
Our findings suggest that uncorrelated noise can promote rather than inhibit coherence in natural systems and that the same effect can be harnessed in engineered systems. \\[2em]
DOI: \href{https://doi.org/10.1103/PhysRevLett.125.094101}{https://doi.org/10.1103/PhysRevLett.125.094101} \\
Phys. Rev. Lett. \textbf{125}, 094101 (2020) 
\end{abstract}

\maketitle

Synchronization, the phenomenon in which oscillators in a population evolve in step with each other, occurs because of interactions or common driving forces among oscillators.  Influences from outside an oscillator network can often be treated as noise, which is usually expected to inhibit synchronization. Indeed, small noise can result in a disproportionately large degree of asynchrony in networks of nonlocally coupled oscillators~\cite{2007_Kawamura_Kuramoto}.  In other contexts, while network disorder can improve synchronous parallel processing performance~\cite{2003_Korniss_Rikvold}, noise has also been found to limit the permissible time delays in communications for network synchronizability and hinders parallel performance~\cite{2010_Hunt}. Still, numerous biological systems---such as neural networks~\cite{2005_Brown_Stopfer, 2003_Yamaguchi_Okamura, 2016_Penn}, ecological communities~\cite{1999_Blasius_Stone}, and the cardiac and cardio-respiratory systems~\cite{1987_Michaels_Jalife, 2013_Kralemann}---and engineered systems---such as arrays of Josephson junctions~\cite{1996_Wiesenfeld_Strogatz}, lasers~\cite{1991_York_Compton}, and nanoelectromechanical devices~\cite{2019_Matheny}---exhibit robust synchronization even under the influence of noise. 

Previous theoretical and experimental observations have demonstrated that \textit{common noise} (in which individual oscillators experience a shared noise term) can actually induce rather than inhibit synchronization~\cite{2002_Zhou_Kurths, 2002_Zhou_Hudson}.  The understanding behind this phenomenon can be traced back to the study of coherence resonance, in which noise leads to greater temporal order in systems with irregular oscillations~\cite{1997_Pikovsky_Kurths,2016_Semenova_Scholl}; to stochastic resonance~\cite{2004_Moss},  
which has been used to reduce the threshold to detect tactile stimuli in human sensory perception~\cite{1996_Collins}; and to the effects of common driving in synchronizing chaotic or disordered systems \cite{1990_Pecora_Carroll,1995_Braiman} as well as the synchronizing effects of periodic driving with a spatially-dependent phase \cite{2006_Brandt}. Synchronization induced by common noise has since been studied in a variety of oscillator networks~\cite{2007_Nakao_Kawamura, 2010_Nagai_Kori, 2016_Pimenova_Pikovsky}.
\begin{figure}
\centering \includegraphics[width=\columnwidth]{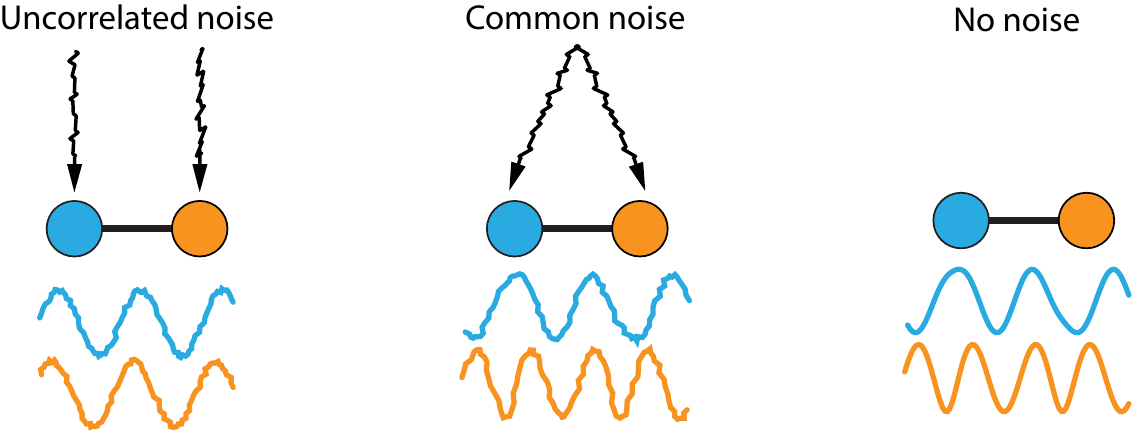}
\caption{Schematics of the main effect. Coupled oscillators experiencing uncorrelated noise exhibit more synchronous dynamics than those subjected to common noise or no noise. This behavior is distinct from those previously observed in oscillator models that synchronize due to large coupling or in response to specific forms of common noise. \label{fig1}}
\end{figure}

Here, we establish the alternative scenario shown in Fig.~\ref{fig1} in which the dynamics are more synchronous in the presence of \textit{uncorrelated noise} than in the absence of noise or even the presence of common noise.  
It seems intuitive that uncorrelated noise would necessarily inhibit synchronization since, unlike the common noise case, it does not have inherent order.  
However, uncorrelated noise is prevalent in many systems, and recent studies suggest the potential for uncorrelated noise to have a positive impact on synchronization. For example, coupled neuronal networks subject to uncorrelated noise 
can exhibit enhanced coherence across the networks while reducing the coherence within each network~\cite{2018_Meng_Riecke}. Uncorrelated noise acting on a pair of oscillators has also been shown to  enhance the phase coherence of one oscillator at the expense of the other~\cite{2015_Amro_Neiman}. Furthermore, uncorrelated noise can promote untwisted phase-locked states over twisted phase-locked states in small-world networks of Kuramoto oscillators~\cite{2012_Esfahani_Samani} and can stabilize an otherwise unstable partially synchronized state in a globally coupled model of oscillators with biharmonic couplings~\cite{2017_Politi_Clusella}. However, the question of whether uncorrelated noise can enhance synchronization to a greater extent than common noise had so far remained open.

We establish our results for several forms of coupled limit-cycle oscillators governed by
\begin{equation}
\frac{{d}\mathbf{x}_i}{{d}t} = \mathbf{f}_i(\mathbf{x}_i) + \frac{K}{N}\sum_{j=1}^N  A_{ij} \mathbf{h}\left(\mathbf{x}_i, \mathbf{x}_j  \right) + \sum_{k=1}^n\mathbf{g}_{ik}(\mathbf{X})\xi_{ik},\label{limit}
\end{equation}
where $N$ is the number of oscillators, $\mathbf{x}_i$ denotes the state of oscillator $i$ (assumed to be $m$ dimensional), $\mathbf{X}=(\mathbf{x}_1, \mathbf{x}_2, \cdots, \mathbf{x}_N)$ encodes the full state of the system, 
$\mathbf{f}_i$ describes the evolution of the isolated oscillators, $\mathbf{h}$ is the coupling function between two oscillators, $K$ is the (tunable) coupling constant, and $A_{ij}$ are the entries of the coupling matrix (assumed to be $1$ if nodes $i$ and $j$ are coupled and $0$ otherwise).
We include $n$ sources of noise determined by the state-dependent direction $\mathbf{g}_{ik}$  and the random variable $\xi_{ik}$, with $\langle \xi_{ik} \rangle =  \langle \xi_{ik}\xi_{jl} \rangle =0$ for all $i,j,k$,and $l\neq k$. The $\xi_{ik}$ term represents multiplicative noise in the case that $\mathbf{g}_{ik}$ varies with $\mathbf{X}$ and represents additive noise in the special case that $\mathbf{g}_{ik}$ is constant.

We assume that in the absence of coupling ($K=0$) and noise ($\xi_{ik}=0$) the isolated node dynamics approach a limit cycle  $\mathbf{x}_i(t) \to \mathbf{x}^c_i(t)$, with $\mathbf{x}_i^c(t) = \mathbf{x}_i^c(t+T_i)$, where $T_i$ is the period of oscillator $i$. We can always define a phase variable $\theta_i(\mathbf{x}_i)$ for oscillator $i$  that, when restricted to the limit cycle, evolves as $\theta_i(t)=\theta_i(0) + \omega_i t$, where $\omega_i=2\pi/T_i$ is the natural frequency. In the presence of coupling, we consider the oscillators to be more synchronized when their relative phase differences are smaller on average. Following Kuramoto~\cite{1975_Kuramoto,2019_Kuramoto_Nakao}, we employ the order parameter $R^2 \equiv  |({1}/{N})\sum_j {e}^{{\mathrm i}\theta_j(t)}  |^2$ as a measure of synchrony, where $\mathrm{i}$ is the imaginary unit. The time-averaged order parameter $\overline{R^2}$ is closer to $1$ when oscillators are more synchronized and closer to $0$ when the oscillators are less synchronized.  In the results below, we say that noise enhances synchronization if $\overline{R^2}$ is larger in the presence of noise than in the absence of noise. We consider two broad forms of noise: common noise, for which $\xi_{ik}=\xi_{jk}$ for all $i$, $j$; and uncorrelated noise, for which $\xi_{ik}$ and $\xi_{jk}$ are independent random variables for all $i\neq j$. We are primarily interested in cases in which uncorrelated noise enhances synchronization more so than common noise. 

\textit{Phase-reduced oscillators.}---For weakly coupled oscillators driven by weak noise, the phase-reduction approximation can be applied to reduce the dynamics of Eq.~\eqref{limit} to a Kuramoto-type model with noise,
\begin{equation}
\label{eom}
\frac{{d}\theta_i}{{d}t} = \omega_i + \frac{K}{N} \sum_j A_{ij} \sin\left(\theta_j - \theta_i\right) + g_i(\theta_i)\eta_i.
\end{equation}
The various noise terms in Eq.~\eqref{limit} result in a single effective noise term $g_i(\theta_i)\eta_i$ in the phase dynamics if, for instance, they are all Gaussian variables with autocorrelations of the same functional form (as shown in Sec.\ S1 of the Supplemental Material~\cite{SM}). The effective noise $\eta_i$ will be assumed to be Gaussian and white unless otherwise noted,
with intensity specified by a matrix  $D_{ij}$ as $\langle \eta_i(t) \eta_j(t') \rangle = D_{ij}\delta\left(t-t'\right)$, where $\delta$ is the Dirac delta function.
The function $g_i(\theta_i)$, called the phase sensitivity function, arises because the effective noise acts on the phase evolution with varying intensity depending on the phase of the oscillator.
In the case of common noise, $\eta_i=\eta_j$ for all $i$, $j$ and all the elements of the noise intensity matrix are identical, with $D_{ij}=\sigma^2/2$ for $\sigma$ denoting the noise intensity. In the case of uncorrelated noise, $\langle \eta_i \eta_j\rangle =0$ for $i\neq j$ and the noise intensity matrix is diagonal, with $D_{ii}=\sigma^2/2$ for all diagonal elements.

\begin{figure*}
\centering\includegraphics[width=2\columnwidth]{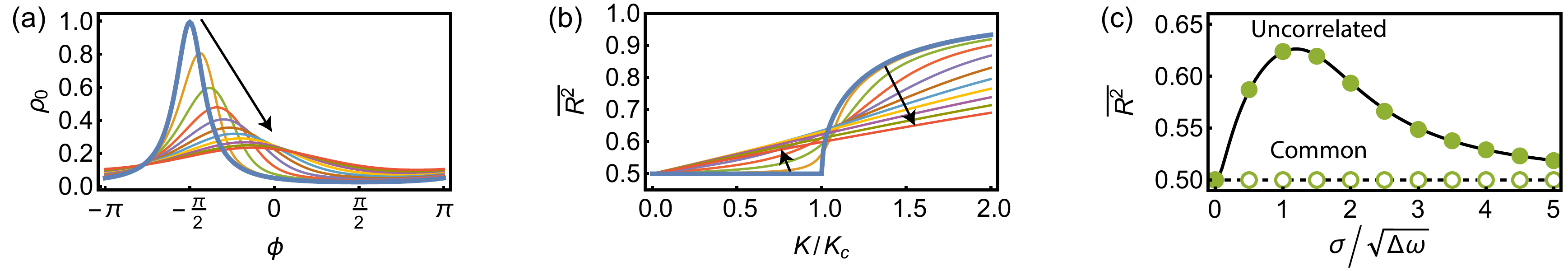}
\caption{Solutions of the Fokker-Planck equation \eqref{fokker} for two phase oscillators subject to Gaussian white noise with constant phase sensitivity $g_i(\theta)=1$.
(a)~Steady ensemble density $\rho_0$ in Eq.~\eqref{exact} as a function of the phase difference $\phi$ for the subcritical coupling $K/K_c = 0.95$.
(b)~Time-averaged order parameter $\overline{R^2}$ as a function of the normalized coupling constant $K/K_c$. The arrows in (a) and (b) indicate the change in the solutions as $\sigma$ increases from zero to $2\sqrt{\Delta\omega}$, where the zero-noise case (thick line) also corresponds to the case of common noise of any intensity. (c)~Time-averaged order parameter $\overline{R^2}$ as a function of the normalized noise intensity $\sigma/\sqrt{\Delta\omega}$ at the coupling constant $K/K_c=0.95$, where the lines show the solutions from the Fokker-Planck equation for uncorrelated noise (continuous) and common noise (dashed). The circles show the agreement with the corresponding direct numerical simulations of Eq.~\eqref{eom}. \label{fig2}}
\end{figure*}

We first consider the case of $N=2$ phase oscillators with $g_i(\theta_i)=1$, so that the multiplicative noise in Eq.~\eqref{limit} becomes additive in the phase approximation.  By moving to a rotating frame, it is possible to take the mean natural frequency equal to zero, so that, without loss of generality, we can take $\omega_1=\Delta\omega/2$ and $\omega_2=-\Delta\omega/2$.  In the absence of noise, the oscillators' phases will drift with respect to each other when the coupling strength $K$ is smaller than $\Delta\omega$, as characterized by their separation angle $\phi \equiv \theta_2-\theta_1$, while their mean angle $\Theta \equiv \left(\theta_1+\theta_2\right)/2$ remains a constant of motion.  They become phase locked as $K$ increases above its critical value $K_c\equiv\Delta\omega$, initially with a separation angle $\phi=-\pi/2$.
In the presence of Gaussian white noise with constant phase sensitivity and for any value of $K$, the evolution of the density of an ensemble of systems $\rho(\phi,\Theta, t)$ can be described by the Fokker-Planck equation
\begin{equation}
\label{fokker}
\frac{\partial \rho}{\partial t} = \frac{\partial}{\partial \phi}\Big[\left(\Delta\omega+ K\sin \phi \right) \rho\Big] + \frac{\sigma^2}{2}\left[\frac{\partial^2 \rho}{\partial \phi^2}+\frac{1}{4}\frac{\partial^2 \rho}{\partial \Theta^2}\right],
\end{equation}
where we have changed variables from $\theta_1$ and $\theta_2$ to $\phi$ and $\Theta$.
Because Eq.~\eqref{fokker} is autonomous with respect to $\Theta$ and $t$, we can find steady solutions which are independent of the mean phase $\Theta$.  Direct integration in this case is possible using an integrating factor. After some simplification, the solution is
\begin{align}
\label{exact}
\rho_0(\phi) \hspace{-0.2em}= & A\int_{0}^{2\pi} {d}\psi~\left[\frac{1}{\exp\left(4\pi\Delta\omega/\sigma^2\right)-1}+H(\phi-\psi)\right] \nonumber \\
& \hspace{-0.7em}\times\hspace{-0.2em}\exp\hspace{-0.2em}\big[2\Delta\omega(\psi\hspace{-0.2em}-\hspace{-0.2em}\phi)/\sigma^2\hspace{-0.2em}-\hspace{-0.2em}2K(\cos \psi\hspace{-0.2em}-\hspace{-0.2em}\cos \phi)/\sigma^2 \big],
\end{align}
for $0\leq \phi \leq 2\pi$, where $A$ is a normalization constant and $H$ is the Heaviside step function.

Figure \ref{fig2}(a) shows how the steady ensemble density $\rho_0$ varies as the noise intensity varies in the case with a subcritical coupling constant $K=0.95K_c$.
The ensemble density, which is peaked near $\phi=-\pi/2$ in the absence of noise, widens and its peak shifts toward zero  as the noise intensity increases.
The widening of the peak represents a loss in one form of coherence, as the oscillator phases become less correlated, but the shifting of the mean difference toward zero represents a gain in a different form of coherence, as the oscillators spend more time with similar phases.
To assess the net impact on synchronization, we consider the time-averaged order parameter, which is determined from the steady-state distribution as $\overline{R^2} = \int_0^{2\pi} \cos^2(\phi/2)\rho_0(\phi){d}\phi$.
Figure \ref{fig2}(b) shows how the time-averaged order parameter varies as the noise intensity varies.  The sharp, phase-locking transition at $K=K_c$ is smoothed out as the noise intensity increases.
For subcritical coupling constants, the order parameter initially increases with increasing noise intensity, indicating an enhancement in synchronization in response to uncorrelated noise that does not occur in the case of common noise. This is illustrated in Fig.~\ref{fig2}(c) for the same $K$ as in Fig.~\ref{fig2}(a), but a qualitatively similar effect occurs for all subcritical cases and for coupling constants just above the critical one.
For large coupling constants, the system is already strongly synchronized in the absence of noise and thus synchronization is not further enhanced by noise.
Time averaging of trajectories from direct numerical simulations of Eq.~\eqref{eom} (see Sec.\ S2 of the Supplemental Material~\cite{SM}) agree extremely well with the solutions derived from the Fokker-Planck equation, as illustrated in Fig.~\ref{fig2}(c). 

\textit{Phase-amplitude oscillators.}---We have shown that phase oscillators can exhibit enhanced synchronization under uncorrelated noise but not under common noise, assuming the noise and coupling terms are weak so that the phase-reduction approximation applies.  We next consider the question of synchronization enhancement in phase-amplitude oscillators experiencing strong noise.
As a prototypical example, we consider $N=2$ coupled Stuart-Landau oscillators each with $m=2$ degrees of freedom $\mathbf{x}_i = \big(x_i^{(1)}, x_i^{(2)}\big)$, which are conveniently represented as a complex variable $z_i(t) = x_i^{(1)}(t) +  {\mathrm i}\, x_i^{(2)}(t)$ and evolve according to
\begin{equation}
\label{stuartlandau}
\frac{{d}z_i}{{d}t} = F_i(z_i) + \frac{K}{4}{\sum_{j=1}^2} (z_i z_j^* - z_j z_i^*)z_i + G_{ik}(z_1,z_2) \xi_{ik}, 
\end{equation}
where $F_i(z_i) \equiv (1+ {\mathrm i}\alpha_i) z_i - (1- {\mathrm i}\gamma_i) \lvert z_i \rvert^2 z_i$ describes the intrinsic dynamics, $\alpha_i$ and $\gamma_i$ are constants, and $*$ denotes complex conjugation. The cubic form of the coupling in Eq.~\eqref{stuartlandau} is selected to result in the Kuramoto-type coupling in the phase reduction, which facilitates comparisons below. 
In the absence of coupling and noise (when $K=0$, $\xi_{ik}=0$), the oscillators have a limit-cycle attractor $z_i(t)=r_i(t){e}^{{\mathrm i}\theta_i(t)}$, where $r_i(t)=1$, $\theta_i(t) = \theta_i(0) + \omega_i t$, and $\omega_i = \alpha_i + \gamma_i$. 

\begin{figure}[t]
\centering\includegraphics[width=\columnwidth]{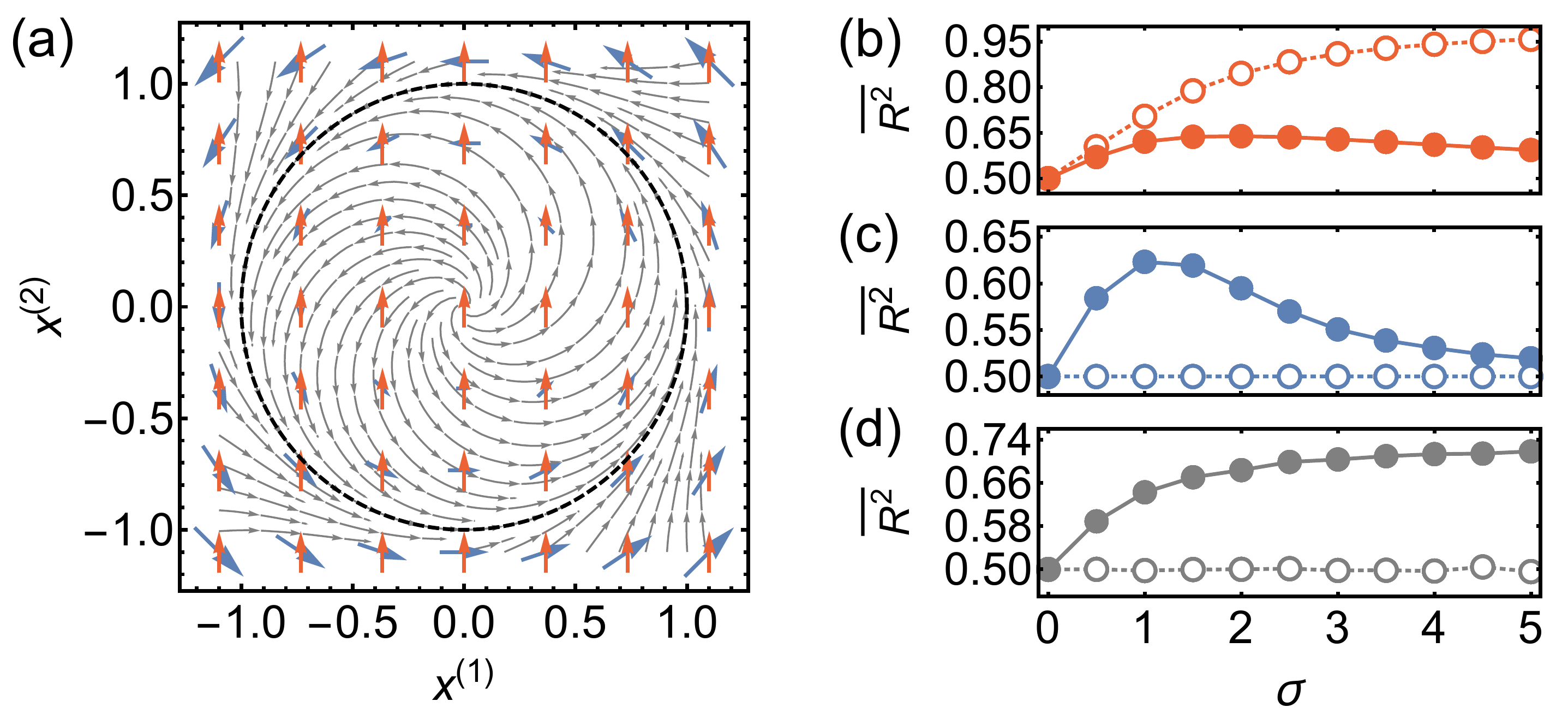}
\caption{Impact of noise on phase-amplitude oscillators. (a)~State space of a Stuart-Landau oscillator, indicating the velocity field in the absence of noise and coupling (continuous lines), the limit-cycle attractor (dashed circle), and the noise forces (arrows). Three forms of noise are represented: Gaussian white noise with $G_{i1}$ (vertical arrows),  Gaussian white noise with $G_{i2}$ (counterclockwise arrows), and Gamma distributed noise with  $G_{i3}$, which acts in the direction of the uncoupled velocity field (continuous lines) when the coupling is small.   
(b)-(d)~Time-averaged order parameter $\overline{R^2}$ as a function of the noise intensity $\sigma$ for correlated (open circles) and uncorrelated (filled circles) noise corresponding to $G_{i1}$ (b), $G_{i2}$ (c), and $G_{i3}$ (d). The oscillator parameters are $\alpha_1=1$ and $\alpha_2=\gamma_1=\gamma_2=0$, and  the coupling constant is $K=0.95$, which is subcritical.
\label{fig3}}
\end{figure}

Figure \ref{fig3}(a) shows the noise forces in the state space of a Stuart-Landau oscillator for three forms of noise determined by differing $G_{ik}$. In each case, the tangent of the noise force along the limit cycle determines the phase sensitivity function in the weak-noise regime for which the phase reduction would hold.  As we proceed with our analysis of the strong-noise regime, it is instructive to compare with predictions for phase-reduced oscillators. 
The phase sensitivity function in the phase reduction for Eq.~\eqref{stuartlandau} takes the form $g_{ik}(\theta_i) = \int_0^{2\pi} d\theta_j \big[G_{ik}(e^{{\mathrm i}\theta_1},e^{{\mathrm i}\theta_2})e^{- {\mathrm i}\theta_i} - G_{ik}^*(e^{{\mathrm i}\theta_1},e^{{\mathrm i}\theta_2})e^{ {\mathrm i}\theta_i}\big]  / 4{\mathrm i} \pi$, where $j\neq i$ \cite{SM}.

Taking additive noise with $G_{i1}(z_1,z_2) = {\mathrm i}$ results in multiplicative noise in the phase reduction with a trigonometric sensitivity function $g_{i1}(\theta_i)=\cos(\theta_i)$, which is expected to induce synchronization under common noise.  
On the other hand, taking $G_{i2}(z_1,z_2) = {\mathrm i}z_i$ results in additive noise in the phase reduction, with a constant sensitivity function $g_{i2}(\theta_i)=1$.
Noise that is modulated by the noiseless part of the dynamics, with $G_{i3}(z_1,z_2) = F_i(z_i)+ ({K}/{4})\sum_j \left(z_i z_j^* - z_j z_i^*\right)z_i$, also results in additive noise in the phase reduction. 
We thus expect, based on our results above for phase oscillators, 
that uncorrelated noise  will enhance synchronization but common noise will not for $G_{i2}$ and $G_{i3}$ also in the unreduced system.

Figure \ref{fig3}(b)-(d) assess these predictions through direct numerical simulations.
For $G_{i1}$, common Gaussian white noise enhances synchronization significantly, as anticipated above,
but uncorrelated noise also enhances synchronization to some extent.  For $G_{i2}$, uncorrelated Gaussian white noise enhances synchronization while common noise does not, which once again agrees with the prediction above. 
To assess if these predictions continue to hold under non-Gaussian noise, we employed noise sampled from Gamma distribution for $G_{i3}$, 
  which is dominated by brief, high-intensity bursts (see Sec.~S2 of the Supplemental Material \cite{SM}). 
It is interesting to note that, while the enhancement for Gaussian cases are qualitatively similar to the phase approximation in Fig.~\ref{fig2}(c), in the non-Gaussian case, the enhancement continues to grow with increasing noise intensity over the same noise range.

In summary, while the synchronization enhancement in Eq.~\eqref{stuartlandau} depends on specific noise features, uncorrelated noise continues to enhance synchronization beyond the phase-reduction approximation in cases where common noise does not. 

\textit{Electrochemical oscillator experiments.}---To test whether the effect described above 
can be observed in real limit-cycle systems, we performed experiments on coupled electrochemical oscillators.
These oscillators, detailed in Supplemental Material Sec.\ S3~\cite{SM} along with sample experimental trajectories, are described by $m=2$ degrees of freedom, which represent the electrode potential and the concentration of the electroactive species in the vicinity of the electrode~\cite{Wickramasinghe_2011}.   Unlike Stuart-Landau oscillators, the limit cycle in this case is not circular in the state space, and thus the phase is a complicated function that we will not attempt to describe analytically. The experimental system consists of two such electrochemical oscillators coupled together through a resistor and the shared fluid environment.

\begin{figure}[b]
\centering\includegraphics[width=\columnwidth]{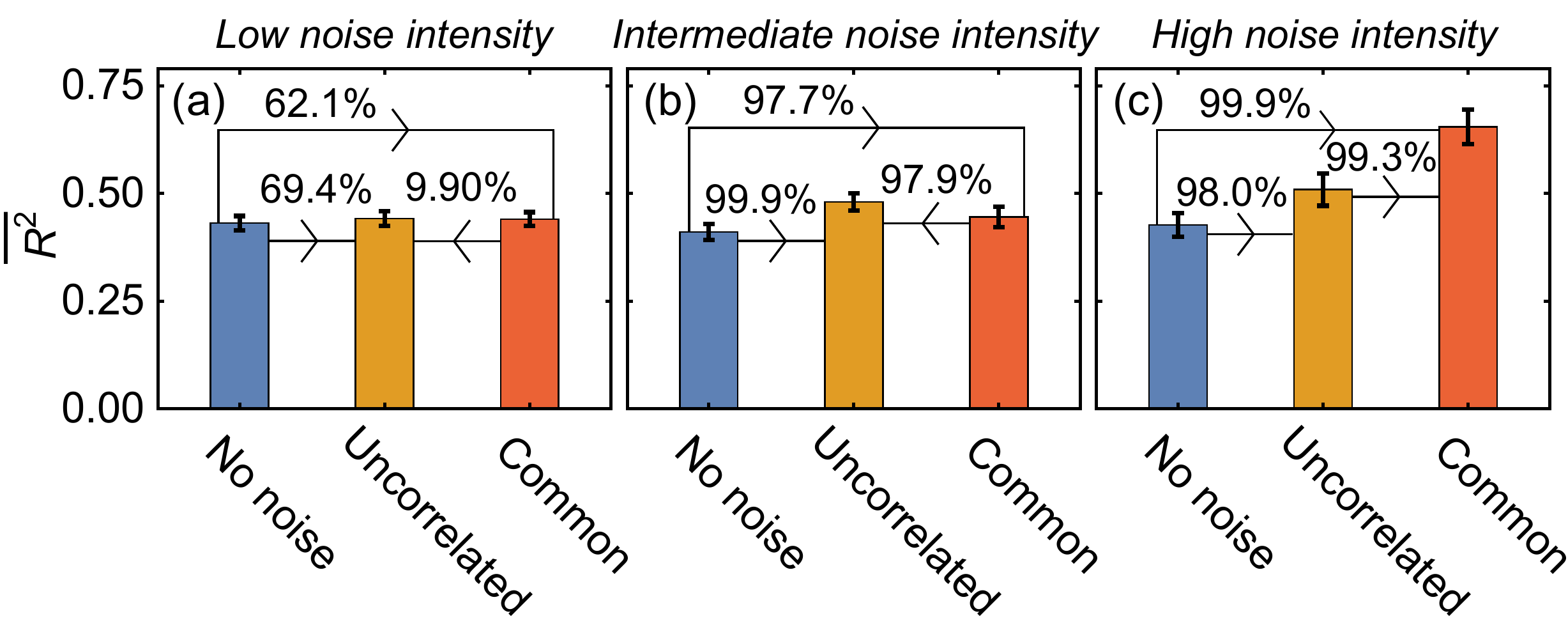}
\caption{Electrochemical oscillator experiments with Gaussian white noise added to the electrode potentials. 
(a)--(c)~Bar plots of $\overline{R^2}$ at three different noise intensities, 
where  $D = 0.025$ V  for $14$ realizations (a),  $D = 0.05$ V for $22$ realizations (b), and  $D = 0.10$ V for $10$ realizations (c). Error bars indicate the estimated errors in the means (i.e., the standard deviation normalized by the square root of the number of realizations), and the arrows between bar plots go from the smaller mean value to the larger mean value with percentages indicating the confidence from a paired $t$ test.  
\label{fig4}}
\end{figure}

For statistical analysis, we create several realizations of the experimental system, with each realization having slightly different natural frequencies and being subject to no noise, common noise, and uncorrelated noise. 
The experiments are repeated for three levels of noise intensity, and the time-averaged order parameter is measured for each experimental run to assess synchronization. 
Figure \ref{fig4} shows the statistical analysis for these experiments. 
We find that for low noise intensity, there is no statistically significant difference between the cases of no noise, uncorrelated noise, and common noise. For intermediate noise intensities,  uncorrelated noise enhances synchronization significantly more so than common noise, confirming the effect described above.  
For high noise intensity, common noise exhibits a greater synchronization enhancement.

We emphasize that, in these experiments, we did not attempt to control the direction of the noise force, given that noise can be easily applied only to the electrode potential and not to the chemical concentration, nor did we attempt to determine (or fine-tune) the phase sensitivity function, given the complexity of the limit cycle.
Nevertheless, we still observe a greater degree of synchronization enhancement for the case of uncorrelated noise than for the case of common noise for intermediate noise intensity.   Thus, these experiments reveal that uncorrelated noise can outperform common noise in synchronization enhancement even without careful design.

\textit{Discussion.}---Our demonstration that uncorrelated noise can enhance synchronization to a greater degree than common noise reveals a new mechanism for how coherent behavior can emerge naturally in spatially-distributed noisy systems. 
The mechanism that generates this noise-enhanced synchronization can be interpreted as follows. On the one hand, when coupled oscillators are close to phase locking, they often spend time at relative angles that are far from zero, and their phases do not add coherently. On the other hand, uncorrelated noise allows the oscillators to escape from these large phase separations and spend more time with similar phases, even when common noise cannot do so precisely because it exerts the same effect on the phases of all oscillators. Our analysis indicates that this effect occurs prominently when the impact of the noise on coupled oscillators is independent of their phases, which means that the coherence is not inherited from a biased filtering of the noise. 

These findings are counterintuitive because the noise terms acting on different oscillators exhibit permutation symmetry for common noise but not for uncorrelated noise; yet, for the systems considered here, the resulting dynamical states are more symmetric in the uncorrelated case. Such synchronization enhancement can thus be interpreted as a manifestation of asymmetry-induced symmetry~\cite{2016_Nishikawa_Motter}, a recently recognized phenomenon in which some degree of asymmetry in a system actually increases the symmetry in the observed state of that system.  

In this study, we observed the preferential enhancement of coherence by uncorrelated noise over common noise in a variety of coupled oscillator systems, including phase and phase-amplitude oscillators, both theoretically and experimentally.    
While we focused here on pairs of oscillators for clarity, we can show that this effect also occurs more generally in larger networks with a frequency gap \cite{2019_DSouza}, such as random networks of Janus oscillators \cite{2019_Nicolaou,2020_Peron} (see Sec.~S4 of the Supplemental Material~\cite{SM}) and multilayer networks relevant to the distributed mammalian neural and circadian systems \cite{Ermentrout_2012,Lilley:2012,Hastings:2018,Reppert:2002} (see Sec.~S5 of the Supplemental Material~\cite{SM}).  
These results overturn the widely held assumption that uncorrelated noise necessarily tampers coherence and suggest that distributed networks that rely on synchronization, such as the network of circadian clocks distributed throughout the body, may benefit from the uncorrelated noise that they experience. 
In contrast with coherence induced by common noise, the enhancement due to uncorrelated noise requires nonvanishing coupling between the oscillators, thus revealing a new relationship between noise and network interactions.

\let\oldaddcontentsline\addcontentsline% Store \addcontentsline
\renewcommand{\addcontentsline}[3]{}% Make \addcontentsline a no-op
\begin{acknowledgments}
The authors acknowledge support from Army Research Office Grant No.~W911NF-19-1-0383 and National Science Foundation Grant No.~CHE-1900011.
\end{acknowledgments}

\let\addcontentsline\oldaddcontentsline
\setcounter{section}{0}
\setcounter{equation}{0}
\setcounter{figure}{0}
\clearpage

\newgeometry{letterpaper,left=1in,right=1in,top=1in,bottom=1in}
\renewcommand{\tiny}{\fontsize{6}{7}\selectfont}
\renewcommand{\scriptsize}{\fontsize{8}{9.5}\selectfont}
\renewcommand{\footnotesize}{\fontsize{9}{11}\selectfont}
\renewcommand{\small}{\fontsize{10}{12}\selectfont}
\renewcommand{\normalsize}{\fontsize{10.95}{13.6}\selectfont}
\renewcommand{\large}{\fontsize{12}{14}\selectfont}
\renewcommand{\Large}{\fontsize{14}{18}\selectfont}
\renewcommand{\LARGE}{\fontsize{17.28}{22}\selectfont}
\renewcommand{\huge}{\fontsize{20.74}{25}\selectfont}
\renewcommand{\Huge}{\fontsize{24.88}{30}\selectfont}
\setlength{\parindent}{15pt}
\renewcommand{\thefigure}{{\small S\arabic{figure}}}
\renewcommand{\thesection}{{\normalsize S\arabic{section}}}
\renewcommand{\theequation}{{\normalsize S\arabic{equation}}}
\normalsize
\makeatletter
\c@secnumdepth=1
\makeatother

\onecolumngrid
\begin{center}
{\Large\bf Supplemental Material for ``Coherent Dynamics Enhanced by Uncorrelated Noise''}\\[0.5em]
{Zachary G. Nicolaou, Michael Sebek, Istv\'{a}n Z. Kiss, and Adilson E. Motter}
\end{center}
\vspace{2em}
\tableofcontents

\section{Effective noise in phase-reduced model}
The model in Eq.~(2) is derived here from the model in Eq.~(1) by adapting the phase reduction technique in Ref.~[30]. To benefit from this technique, we assume that $\xi_{ik}$ 
and the coupling term are all small and can be accounted for by their leading order contributions.  
Because these quantities are small, we can also assume that the time evolution remains close to the limit cycle.  The phase $\theta_i$ is defined as a function of the state $\mathbf{x}_i$   such that  $\nabla_{\mathrm{x}}\theta_i(\mathbf{x}^c_i) \cdot \mathbf{f}_i(\mathbf{x}^c_i) = \omega_i$, ensuring that the phase increases linearly with the natural frequency on the limit cycle when the oscillators are uncoupled. In the presence of coupling and noise, the evolution of the phase is then given by
\begin{equation}
\frac{{d}\theta_i}{{d}t} = \omega_i + \nabla_{\mathrm{x}}\theta_i \cdot \Big[ \frac{K}{N}\sum_{j=1}^NA_{ij}\mathbf{h}(\mathbf{x}_i,\mathbf{x}_j) + \sum_{k=1}^n \mathbf{g}_{ik}(\mathbf{X})\xi_{ik} \Big],
\end{equation}
where we used that ${d}\theta_i/{d}t =  \nabla_{\mathrm{x}}\theta_i \cdot {{{d}\mathbf{x}_i}/{d}t}$.

Taking the leading order approximation, the gradient $\nabla_{\mathrm{x}}\theta_i$ is replaced with the value at the corresponding point on the limit cycle, which we denote by $\mathbf{Z}_i(\theta_i)$. Noting that the difference $\theta_i - \omega_i t$ is a slow variable, its evolution can be determined by calculating time averages of the fast variables, which are approximated by averages over a single cycle.
This leads to
\begin{equation}
\frac{{d}\theta_i}{{d}t} = \omega_i + \frac{K}{N}\sum_{j=1}^N A_{ij} \Gamma_i(\theta_i - \theta_j) + \sum_{k=1}^n g_{ik}(\theta_i)\xi_{ik} ,
\end{equation}
where the coupling function and phase sensitivity are 
\begin{align}
\Gamma_i(\theta_i - \theta_j) &=\int_{0}^{2\pi} \mathbf{Z}_i(\theta_i + \phi )\cdot\mathbf{h}\big[\mathbf{x}_i^c(\theta_i + \phi),\mathbf{x}_j^c(\theta_j + \phi)\big] \frac{{d}\phi}{2\pi},  \\
g_{ik}(\theta_i) &= \int_0^{2\pi} \int_0^{2\pi} \cdots \int_0^{2\pi} \mathbf{Z}_i(\theta_i )\cdot\mathbf{g}_{ik}\big[\mathbf{X}^c(\theta_1, \theta_2, \cdots, \theta_N)\big]\prod_{j\neq i}\frac{{d}\theta_j}{2\pi},
\end{align}
respectively, with $\mathbf{X}^c(\theta_1, \theta_2, \cdots, \theta_N) = (\mathbf{x}_1^c(\theta_1), \mathbf{x}_2^c(\theta_2), \cdots, \mathbf{x}_N^c(\theta_N))$. 
It is desirable to define an effective noise $\eta_i$ with phase sensitivity $g_i(\theta_i)$ such that the random variables $g_i(\theta_i)\eta_i$ and $\sum_k g_{ik}(\theta_i)\xi_{ik}$ are drawn from the same distributions for all $\theta_i$.  In the general case, the temporal correlations in the random variable $\xi_{ik}$ 
cannot be entirely factored from the phase dependencies in $g_{ik}(\theta_i)$.  However, if the $\xi_{ik}$  are independent for differing $k$ and are stationary zero-mean Gaussian processes with the same autocorrelation functions (so that $\langle \xi_{ik}(t) \rangle =  0$,  $\langle \xi_{ik}(t_1)\xi_{jl}(t_2) \rangle = 0$ for all $i,j,k$, and $l\neq k$, and $\langle \xi_{ik}(t_1)\xi_{ik}(t_2) \rangle = C_i(t_1-t_2)$ for all $k$), then the sum of Gaussian variables is again Gaussian and the mean and autocorrelations completely specify the resulting distribution for each $\theta_i$.  This leads to an effective phase sensitivity $g_i(\theta_i) = \left[\sum_k g_{ik}(\theta_i)^2\right]^{1/2}$ and $\theta_i$-independent Gaussian $\eta_i(t)$ with the same autocorrelation $C_i$.
For the Kuramoto-type  coupling function $\Gamma_i(\theta_i-\theta_j) = \sin(\theta_j-\theta_i)$, the reduction dynamics correspond to Eq.~(2). 

In studies of synchronization induced by common noise, the most frequently employed phase sensitivity has been trigonometric, as in $g_i(\theta_i)=\cos(\theta_i)$.  A different but important case considered in the main text is the constant phase sensitivity function, $g_i(\theta_i)=1$, corresponding to additive noise in the phase reduction. This choice is natural in systems that do not have a preferred phase, so that the dynamics are invariant under global phase rotations.  In the case of $n=1$ noise terms, $\mathbf{g}_{i1}(\mathbf{X}^c) = \mathbf{Z}_i(\theta_i)/\lVert\mathbf{Z}_i(\theta_i)\rVert^2$ results in $g_i(\theta_i)=1$, which, for example, follows after rescaling the noise intensity for any rotationally-invariant noise in the Stuart-Landau equation or for $\mathbf{g}_{i1}(\mathbf{X})$ proportional to the noiseless part of the dynamics more generally. For common noise with $\eta_i(t) = \eta(t)$ for all $i$, this rotational symmetry permits a change in variables $\theta_i \to \theta_i +  \int_0^t {d}t' \eta(t')$, which eliminates the noise from Eq.~(1).  It follows that under common noise the order parameter must be identical to that for the case without noise, and thus common noise cannot affect synchronization in the case of constant phase sensitivity, whereas, as evidenced by Fig.~2, uncorrelated noise enhances synchronization.  Similar results are expected more generally for noise that results in approximately constant $g_i(\theta_i)$. 

\section{Direct numerical integration and calculation of $\overline{R^2}$}
\begin{figure}[t]
\includegraphics[width=0.9\columnwidth]{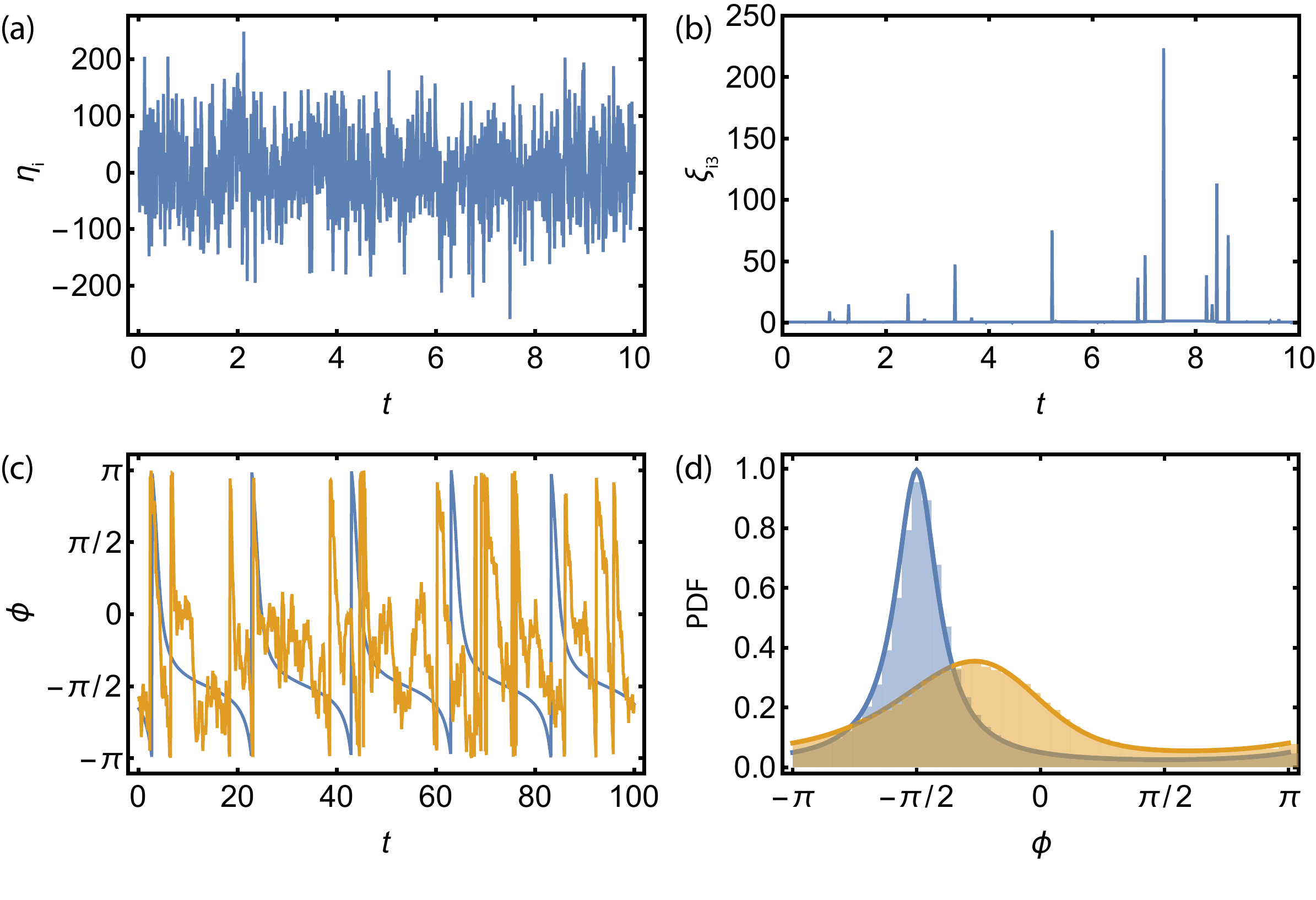}
\caption{Direct numerical simulations for Eqs.~(2) and (5). (a) Gaussian white noise $\eta_i$ used in simulation of Eq.~(2) vs.\ $t$, sampled with a rate $\tau=10^{-4}$. Such noise was also used for $\xi_{i1}$ and $\xi_{i2}$ in simulations of Eq.~(5). (b) Gamma-distributed noise $\xi_{i3}$ used in simulations of Eq.~(5) vs.\ $t$, sampled with a rate $\tau=10^{-2}$. (c) Phase difference $\phi=\theta_2-\theta_1$ vs. time for sample simulations corresponding to Fig.~(2) of the main text, with $g_i(\theta)=1$ for common noise (blue) and uncorrelated noise (orange). (d) Empirical probability distribution function (PDF) for time spent at each phase separation $\phi$ derived from the trajectories in (c), with reference lines showing the exact solution $\rho_0$  of the Fokker-Planck equation in Eq.~(4). \label{fig_sim}}
\end{figure}

Direct numerical integration of Eqs.~(2) and (5) was carried out in the sense of It\^{o} using the GNU Scientific Library. Since computations necessarily use finite time steps, it is not possible to simulate true white noise, which varies on arbitrarily short time scales. Instead, noise is sampled from a Gaussian distribution at a fixed sampling rate of $1/\tau$. The variance of the sampled noise is normalized by $1/\tau$ in order for the $\tau \to 0$ limit to correspond to Gaussian white noise. For the non-Gaussian case of Eq.~(5) with $G_{i3}$, we instead sample $1+\xi_{i3}$ from a Gamma distribution with mean one and variance $\sigma^2/\tau$. In the latter case, noise is dominated by short bursts with large amplitude, which resembles pulse-like noise relevant to the study of neurons~[37]. Figure \ref{fig_sim}(a) shows a sample of Gaussian white noise sampled with a rate $\tau=10^{-4}$, which is sufficiently small to approximate white noise well, while Fig.~\ref{fig_sim}(b) shows the Gamma-distributed noise sampled with $\tau=10^{-2}$.

Figure \ref{fig_sim}(c) shows example trajectories of the system in Eq.~(2) with common noise and uncorrelated noise in the case of constant phase sensitivity $g_i(\theta)=1$.  Since the attractors in the system are ergodic, the empirical distribution of time the oscillators spend with relative phase separation $\phi$ is given by the solution of the Fokker-Planck equation, as shown in Fig.~\ref{fig_sim}(d).  Accordingly, the order parameter $\overline{R^2}$ is computed from  time averages of $\cos^2(\phi/2)$ over the trajectories, which, as shown by the dots in Fig.~2(c), agrees extremely well with the exact Fokker-Planck solution.  
Note than the average value $\overline{R}$, on the other hand, would be computed from time averages of the non-analytic function $\lvert \cos(\phi/2)\rvert$.
Consequently, while the expectation for the subcritical noiseless case ($\sigma=0$ and $K\leq K_c$) with constant phase sensitivity is $\overline{R^2}\equiv 1/2$ identically, the expectation $\overline{R}$ varies with $K$. Thus, $\overline{R^2}$ is a more proper order parameter in the case of two phase oscillators than $\overline{R}$, since it more clearly delineates the noiseless phase-locking transition. Nevertheless, $\overline{R}$ and $\overline{R^2}$ exhibit the same trends with varying $\sigma$, so synchronization is enhanced by uncorrelated noise according to either metric. 

In summary, the results from direct numerical simulations in Fig.~\ref{fig_sim}(c)-(d) completely agree with those from the Fokker-Planck equation presented in Fig.~2 of the main text. The oscillators spend more time at small phase separations $\phi$ for uncorrelated noise than for common noise, leading to enhanced synchronization.

\section{Electrochemical oscillator experiments}
\begin{figure}[t]
\centering\includegraphics[width=0.9\columnwidth]{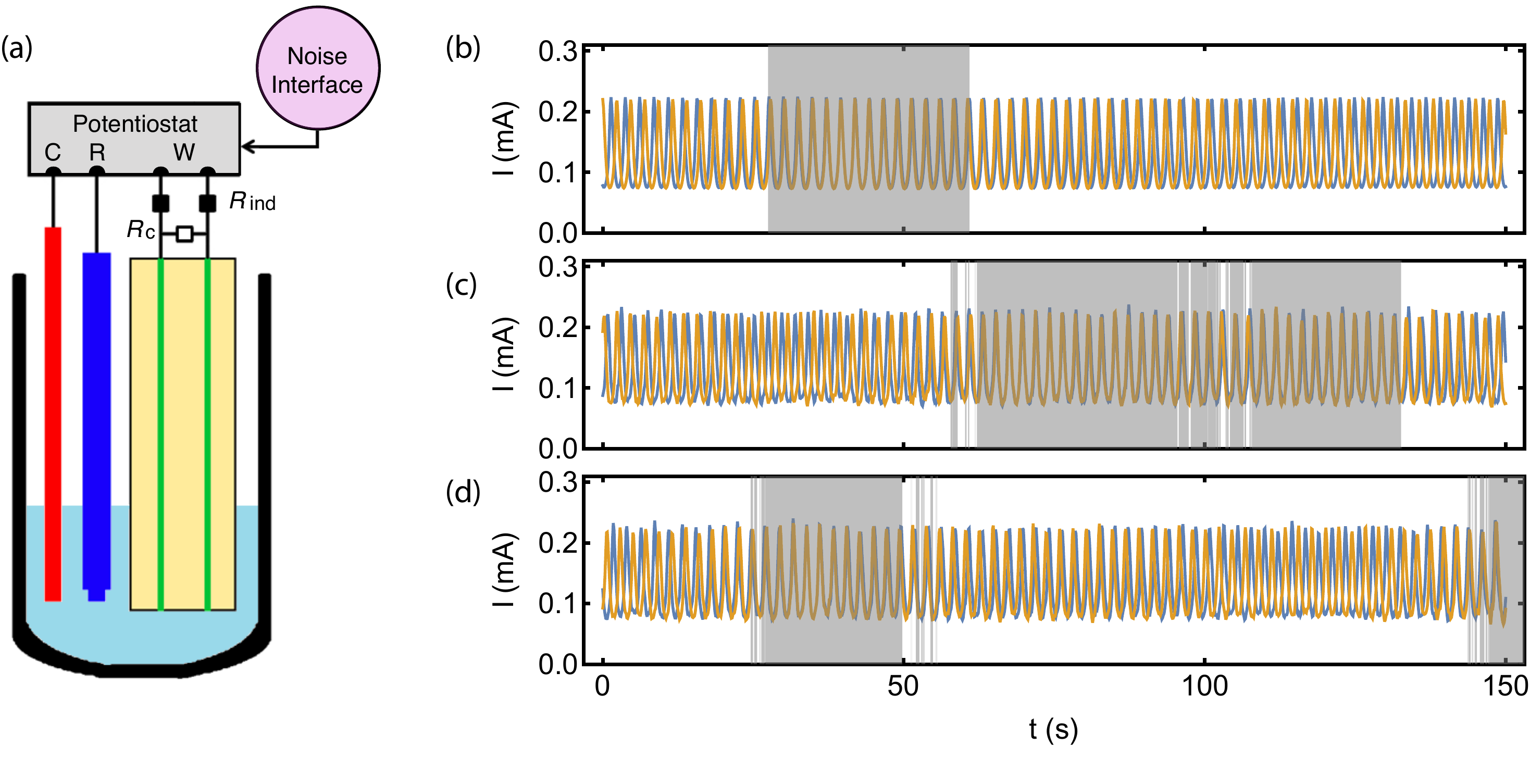}
\caption{Electrochemical oscillator experiments. (a) Schematic of the experimental setup, showing the electrochemical cell with a counter electrode C, a reference electrode R, and two working electrodes W connected through $R_{\mathrm{ind}}=1$ k$\Omega$ resistors. The electrodes are coupled through a cross resistance $R_c$, which gives rise to a coupling strength of $K=1/R_c$, and noise is applied through an interface that superimposes an additional potential to the electrode. (b)-(d) Measured currents $I$ vs.\ time $t$ for the first (blue lines) and second (orange lines) oscillators, with no noise (b), uncorrelated noise (c), and common noise (d), where the gray highlights periods for which the magnitude of the phase difference was less than $\pi/4$. 
 \label{figS1}}
\end{figure}

An electrochemical cell was~built with a platinum coated titanium rod counter electrode,
Hg/Hg\textsubscript{2}SO\textsubscript{4}/(sat.)~K\textsubscript{2}SO\textsubscript{4} reference electrode, 25 working nickel electrodes embedded in epoxy (each 1 mm in diameter), and 3 M H\textsubscript{2}SO\textsubscript{4} held at
10${^o}$C, as shown schematically in Fig.~\ref{figS1}(a). A multi-channel potentiostat was used to apply constant potential
($V\textsubscript{0}$) to each electrode through individual resistors ($R\textsubscript{ind}$) of 1 k$\Omega$, and the
oscillatory current was measured at a rate of 200 Hz. Noise was injected into the system employing a feedback interface to modulate the potential on the $i${th} electrode, $V_{i}(t) = V_{0} + \eta_{i}(t)$,
 where $\eta_{i}$ is Gaussian white noise generated using Matlab.
The natural frequency of each electrode was determined without noise. Coupling was added through an external resistance
between two electrodes, 
which provided multiple pairs of electrodes exhibiting
phase slipping behavior near the onset of synchrony. A LabView program was created to switch between uncorrelated noise, no noise, and common noise as data was collected. 

Traces of the experimentally measure current $I$ are shown in Fig.~\ref{figS1}(b)-(d) for realization with no noise, uncorrelated noise, and common noise.  Oscillators have natural frequencies around $0.5$~Hz and, in the absence of noise, exhibit regular periods of near synchrony separated by phase slips. 
As is the case for phase oscillators with constant phase sensitivity function, uncorrelated noise has the effect of lengthening the periods of time the oscillators can spend with small phase separations, as shown by the highlighted areas in Fig.~\ref{figS1}(b)-(d).  Experimental runs were repeated to yield multiple datasets for analysis at weak noise ($D$ = 0.025 V, 14 pairs), intermediate noise ($D$ = 0.050 V, 22 pairs), and strong noise ($D$ = 0.10 V, 10 pairs).   Since the experiments have limited length of time, the averaging over multiple experimental realizations shown in Fig.~4 is necessary to establish this effect statistically.

\section{Random networks of Janus oscillators}
\begin{figure}[t]
\includegraphics[width=0.9\columnwidth]{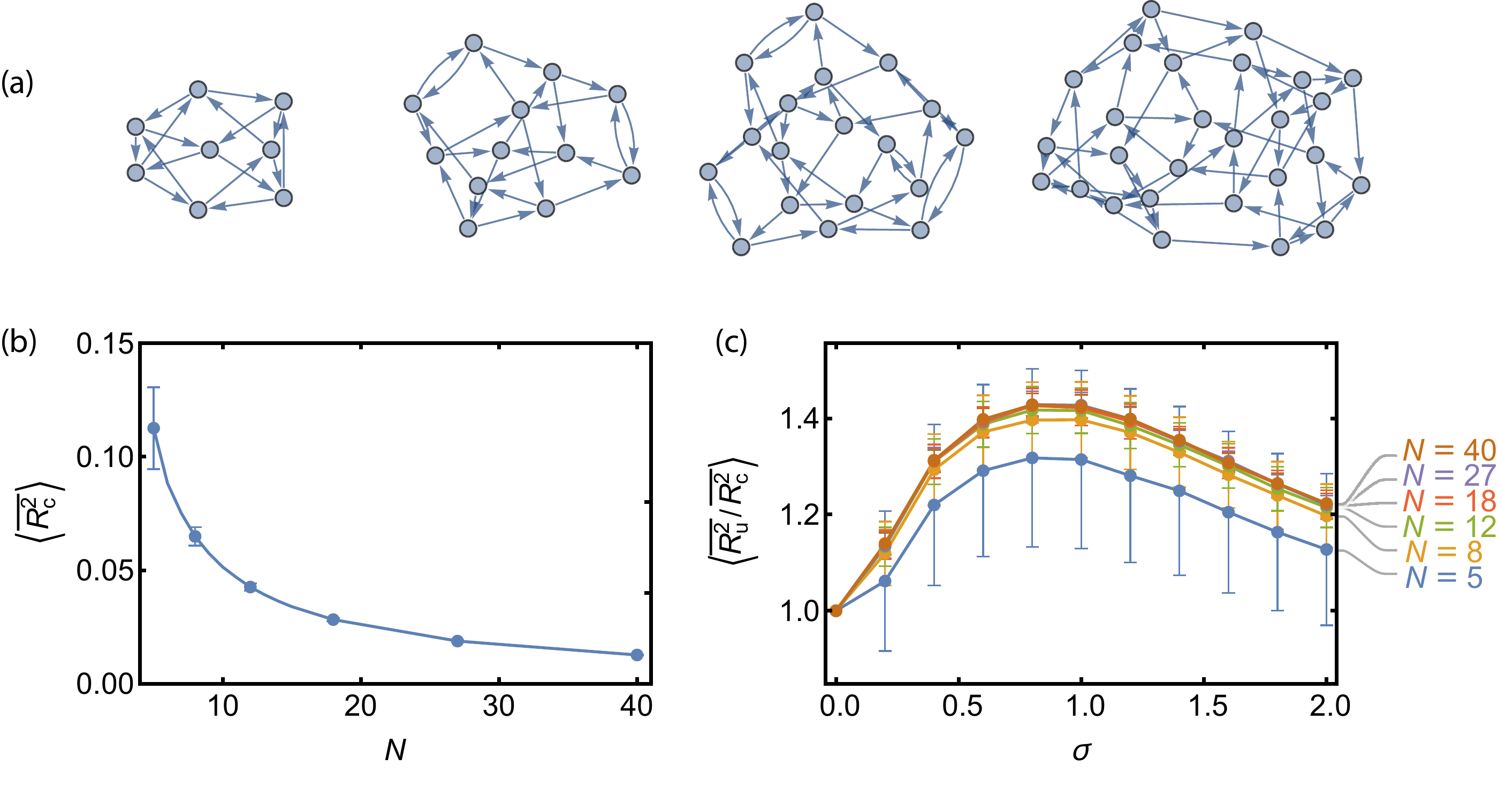}
\caption{Effects of noise in Janus oscillator networks. (a) Random networks of Janus oscillators, with arrows from oscillator $j$ to oscillator $i$ indicating coupling from $\theta_j^2$ to $\theta_i^1$. (b) Network average of the order parameter $\langle\overline{R_c^2}\rangle$ over $100$ random network realizations vs.\ number of Janus oscillators $N$ for oscillators subject to common noise, with $\Delta\omega=1$ and $K=\beta=1/6$ and error bars showing the standard deviation over the random network realizations. (c) Network average of the order parameter for  uncorrelated noise, $\overline{R_u^2}$, relative to that for common noise, $\overline{R_c^2}$, vs.\ noise intensity $\sigma$ for random networks of various sizes, with error bars as in (b). \label{fig_janus}}
\end{figure}
While we focused on the case of $N=2$ phase oscillators in the main text,  
uncorrelated noise can also enhance synchronization when common noise does not in classes of random networks of oscillators.  
For concreteness, we illustrate this in a networks of coupled Janus oscillators, which have been of recent interest for their potential to produce rich dynamics~[35, 36]. Each Janus oscillator is composed of two phase-oscillator components with opposite natural frequencies, and pairs of Janus oscillators interact through a coupling term between phase-oscillator components of opposite natural frequencies.  
Networks of Janus oscillators can also be regarded as multilayer networks consisting of two layers of phase oscillators, each with a different natural frequency. We consider the impact of common and uncorrelated noise acting on each phase-oscillator component, which evolves according to
\begin{align}
\frac{d\theta_i^1}{dt} &= \frac{\Delta \omega}{2} + \beta \sin(\theta_i^2-\theta_i^1) + K \sum_j A_{ij} \sin(\theta_j^2-\theta_i^1) +  \eta_i^1, \label{janus1} \\
\frac{d\theta_i^2}{dt} &= -\frac{\Delta \omega}{2} + \beta \sin(\theta_i^1-\theta_i^2) + K \sum_j A_{ji} \sin(\theta_j^1-\theta_i^2) +  \eta_i^2, \label{janus2}
\end{align}
where $\Delta \omega$ is the frequency gap, $\beta$ and $K$ are the internal and external coupling constants, respectively, $A_{ij}$ are the components of the adjacency matrix, and $\eta_i^k$ are Gaussian white noise terms with intensity $\sigma$. While here we focus on Eqs.~\eqref{janus1}-\eqref{janus2} for concreteness, we expect similar conclusions to hold in wider classes of multilayer and multiplex networks, which have been of recent interest in the study explosive phenomena~[34]. In contrast to the globally-coupled Kuramoto model, there are relationships between the frequency distribution and network structure in these models, which can facilitate synchronization enhancement by uncorrelated noise.

Figure \ref{fig_janus}(a) shows a selection of random regular networks of Janus oscillators, in which  each oscillator has both in- and out-degree equal to $2$.
The noise is either common (with $\eta_i^1=\eta_j^2$ for all $i,j$) or uncorrelated (with $\langle \eta_i^1 \eta_j^2 \rangle = \langle \eta_i^1 \eta_j^1 \rangle = \langle \eta_i^2 \eta_j^2 \rangle =0$ for all $i,j$).  Because the degrees of all nodes are identical, all oscillators are close to phase locking with their neighbors for slightly subcritical $K$, as it was the case for the systems of $N=2$ oscillators considered in the main text.  Since common noise can be eliminated in a rotating reference frame, the average order parameter in the presence of common noise $\overline{R_c^2}$ is equal to the noiseless case for each network realization. The average of the order parameter over the network realizations, $\langle\overline{R_c^2}\rangle$, decreases with increasing network size, as shown in Fig.~\ref{fig_janus}(b). The impact of uncorrelated noise should thus be measured relative to that of common noise to compare networks of differing sizes, as shown in Fig.~\ref{fig_janus}(c).  Regardless of network size, uncorrelated noise enhances synchronization for intermediate noise intensity by allowing oscillators to spend less time at large phase separations.  Thus, the constructive effect of uncorrelated noise described in the main text generalize beyond two oscillators to classes of random networks.

\section{Layered networks of phase oscillators}
Networks of oscillators in biological systems often have layered structure, and the impact of uncorrelated noise on such networks is of broad interest. Here, we consider a two-layer version of the system in Eq.~(2) of the main text. 
 The first layer consists of $N-2$ globally-coupled phase oscillators,
\begin{equation}
\label{eom2}
\frac{{d}\theta_i}{{d}t} =
\omega_i + \frac{K'}{N-2} \sum_{j=1}^{N-2} \sin\left(\theta_j - \theta_i\right),\,  i=1,\cdots, N-2,
\end{equation}
where $K'=K''\left[1+\cos\left({\theta_{N}-\theta_{N-1}}\right)\right]/2$ is a coupling strength that varies with the phases of the oscillators in the second layer and $K''$ is a tunable constant. Oscillator heterogeneity is included in this model by sampling the natural frequencies from a Lorentz distribution centered at zero with width $\gamma=1$.
The second layer consists of two coupled oscillators that are subject to Gaussian white noise,
\begin{equation}
\label{eom3}
\frac{{d}\theta_i}{{d}t} =
\omega_i + \frac{K}{2} \sum_{j=N-1}^N \sin\left(\theta_j - \theta_i\right) + g_i(\theta_i)\eta_i,\, i=N-1,N,
\end{equation}
where $K$ is a constant. 

The model in Eqs.~\eqref{eom2}-\eqref{eom3} represents scenarios in which coupling between oscillators depends on time-dependent resources, described here by the variable coupling strength $K'$. For instance, certain properties of neurons that impact their coupling strengths are modulated by the presence of circadian-regulated hormones such as cortisol~[38].
The circadian system relies on the synchronization of multiple interacting clocks located both in the brain~[39] and in tissues outside the brain, including the liver and kidneys~[40]. These clocks can experience common and uncorrelated noise owing to the differential impact of the various environmental signals they receive, which include light, available nutrients, sleep schedule, and tissue-dependent conditions. In this example, $\theta_{N-1}$ and $\theta_N$ in Eq.~\eqref{eom3} would represent the phases of circadian clocks whereas $\theta_1, \ldots, \theta_{N-2}$ in Eq.~\eqref{eom2} correspond to the phases of neurons whose coupling strength is modulated by the circadian-regulated hormones.

\begin{figure}[t]
\centering\includegraphics[width=0.9\columnwidth]{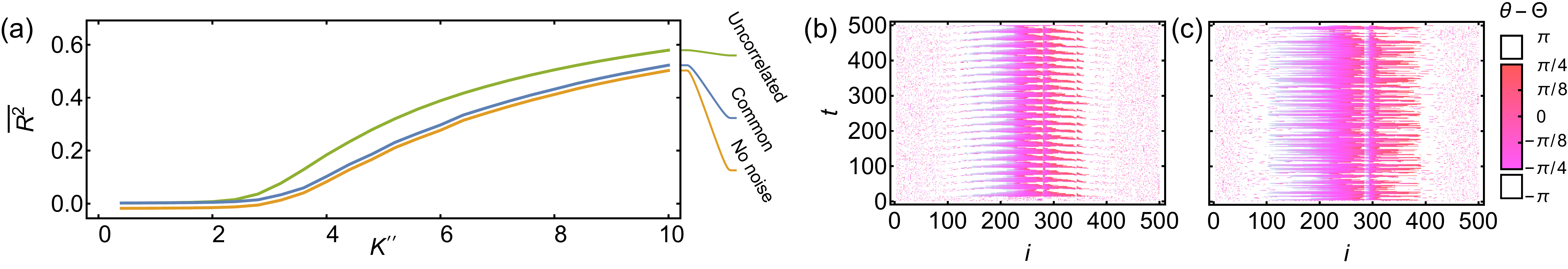}
\caption{Synchronization enhanced by uncorrelated noise in a two-layer network. (a) Time-averaged order parameter $\overline{R^2}$ as a function of the coupling parameter $K''$ for $N-2=500$ oscillators, averaged over $10$ frequency realizations. For the driving layer, the two-oscillator coupling strength is $K=0.95$, the natural frequencies are $\pm 0.5$, and the noise intensity is $\sigma=1.0$. (b)-(c) Example realization of the time evolution of the oscillator phases relative to the mean phase for common noise (b) and uncorrelated noise (c) with $K''=5.0$ and the other parameters as in (a).  To facilitate visualization, oscillators with phases far from the mean phase are colored white, as indicated in the color bar.
\label{figS2}}
\end{figure}

We analyze the response of the time-averaged order parameter $\overline{R^2}$ to common noise, uncorrelated noise, and no noise, as shown in Fig.~\ref{figS2}(a) for a constant function $g_i(\theta_i)=1$.
For all coupling constants considered, the order parameter is larger in the case of uncorrelated noise than in the cases of common noise and no noise, where the latter two remain comparable.
The enhancement in synchronization can also be appreciated visually from the phase evolution of individual oscillators, as shown in Fig.~\ref{figS2}(b)-(c).
Therefore, it is clear that uncorrelated noise can enhance synchronization in large networks when common noise does not, and this effect may play a role in maintaining the synchrony in biological systems. 

\end{document}